# On-chip optical filters with designable characteristics based on an interferometer with embedded silicon photonic structures

Serdar Kocaman[1*], Mehmet S. Aras[1], Nicolae C. Panoiu[2], Ming Lu[3], and Chee Wei Wong[1*]

[1]*Optical Nanostructures Laboratory, Columbia University, New York, NY 10027, USA*
[2]*Department of Electronic and Electrical Engineering, University College London, London WC1E 7JE,UK*
[3]*Center for Functional Nanomaterials, Brookhaven National Laboratory, Upton, NY 11973, U.S.A.*
**Corresponding author: sk2927@columbia.edu ; cww2104@columbia.edu*

We demonstrate chip-scale flat-top filters at near-infrared wavelengths using negative index photonic crystal based Mach Zehnder interferometers. Supported by full three-dimensional numerical simulations, we experimentally demonstrate a new approach for engineering high-pass, low-pass, band-pass, and band-reject filters, based on designing the photonic band diagram both within the band-gap frequency region and away from it. We further show that our approach can be used to design filters that have tunable multi-level response for different sections of the spectrum and for different polarizations. This configuration enables deterministic control of the bandwidth and the rejection ratio of filters for integrated photonic circuits.

Dielectric structures which are periodic in one dimension have been used as frequency filters for electromagnetic (EM) radiation since decades ago [1]. With the development of photonic crystals (PhCs) in late 1980s [2, 3], structures with wavelength-scale periodicity in two and three dimensions enabled the study of many new phenomena [4-6]. One particular example of periodically structured materials is left-handed metamaterials (LHMs) [7-9]. In most cases, LHMs are made of metallic resonators, which at optical frequencies [9] can lead to large losses. To overcome this shortcoming, one can employ PhCs made of dielectrics, which can have effective negative index of refraction in certain frequency ranges [10-11] as well as low optical losses. Important applications of this class of LHMs have been recently demonstrated, including zero phase delay lines and sub-diffraction imaging [12-14].

In this letter, we introduce on-chip PhC-based designs of optical filters, which operate at wavelengths at which photonic bands exist (the material is transparent), unlike most of the commonly used PhC-based optical filters, which operate chiefly at or near the bandgap wavelengths [15-17]. We demonstrate that by incorporating negative index PhCs in a Mach Zehnder interferometer (MZI) a new class of optical filters can be achieved, namely filters based on the phase effects arising from the negative refractive index characteristic of PhCs. More specifically, unlike most free-space interferometer applications, the phase difference that leads to interference in the proposed devices comes from the imbalance in the refractive indices of the two arms of the MZI and not the physical length difference. By combining in the same design PhCs with negative and positive index of refraction, one can broadly engineer the corresponding difference in the index of refraction. This approach is made possible by the availability of precise fabrication techniques in the integrated photonics, and has already been used by several studies [12, 18].

The hexagonal PhC regions [Fig. 1(a) left inset] inserted in the two arms of the MZI are made of air holes etched into a silicon (Si) slab ($n_{Si}$=3.48), with lattice period $a$ = 430 nm and slab thickness $t$ = 320 nm, placed on top of a silica substrate ($n_{SiO2}$ = 1.46). The PhC band diagram with a hole-to-lattice ratio $r/a$=0.285 (radius $r$ = 122 nm) is shown in Fig. 1(b)-(c). The two-dimensional (2D) PhC has a negative refractive index in the spectral range 0.3-0.273, in normalized frequency of $\omega a/2\pi c$, or 1433-1575 nm in wavelength [12].

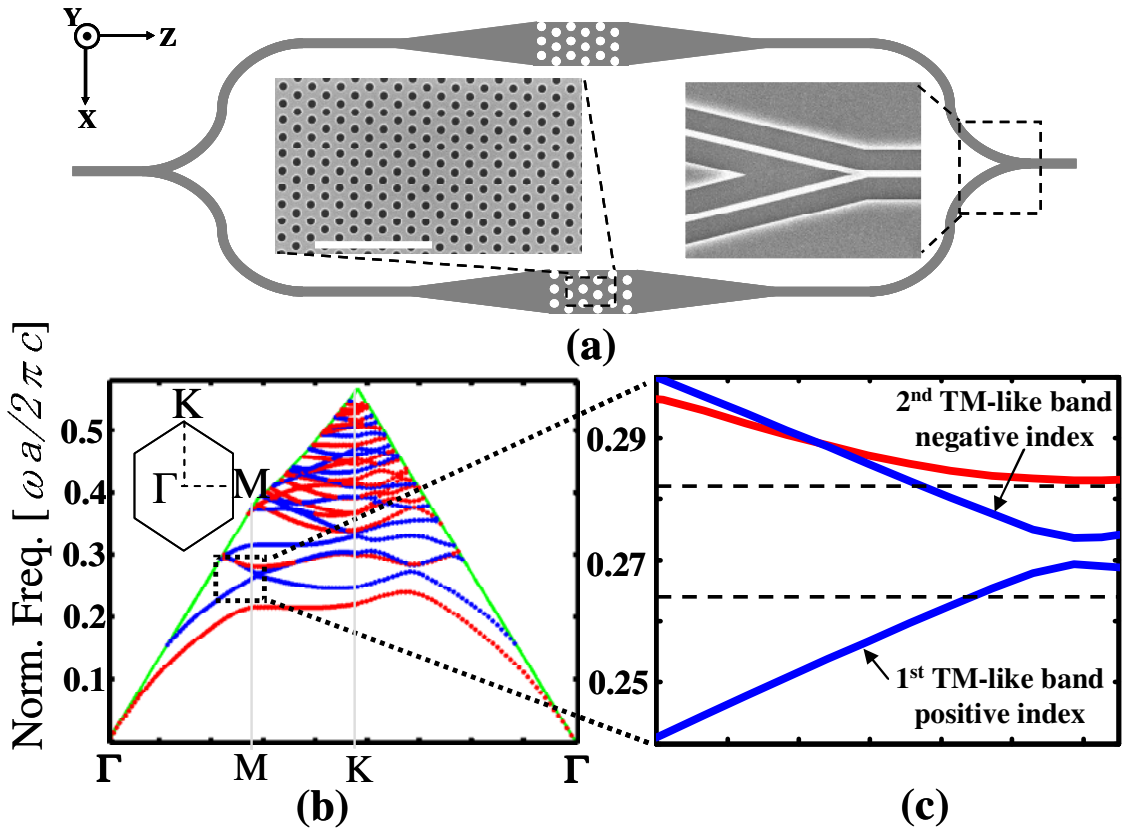

**Fig 1.** (a) Schematic representation of the MZI. (b) Band diagram of the Inset: first Brillouin zones of the hexagonal PhC. The TM-like (TE-like) photonic bands are depicted in blue (darker) [red (lighter)]. (c) A zoom-in of the spectral domain corresponding to experimental region of interest. Experiments were performed in the spectral region marked by the two horizontal dashed lines.

We have proceeded with the design of the device by calculating the band structure of the PhC (Fig. 1b). 30 bands and the effective refractive index for each band have been calculated for the full 3D case. The photonic bands are classified as TE-like or TM-like according to how the modes transform on applying the reflection operator, $\sigma_y$, defined with respect to the *xz* plane; the parameter $\sigma_y = \int_{BZ} E_k(\mathbf{r}) \Sigma_y E_k(\mathbf{r}) d\mathbf{r}$ quantifies the mode parity. The computed values of $\sigma_y$ (~ ± 0.98) are used to distinguish between the even ($\sigma_y$>0, TE-like) and odd ($\sigma_y$<0, TM-like) modes. The effective refractive indices are determined from the relation $k = \omega |n| / c$, where $k$ is the wavevector in the first Brillouin zone. Note that since $k$ decreases with $\omega$ for the second TM-like band in Fig. 1(c), the effective index of refraction is negative [13]. Here, the design parameters for the band diagram are the thickness of the slab, *t*, the radius of the holes, *r*, and the period of the crystal lattice, *a*.

In order to characterize the sensitivity of the optical properties of the PhC to changes of the crystal parameters we have varied *t*, *r* and *a*, one at a time, and determined the corresponding variations of the band diagram. Results of three simulations for each case are summarized in the Fig. 2. Since we designed our filters to operate for TM polarized waves we only show here the TM-like bands.

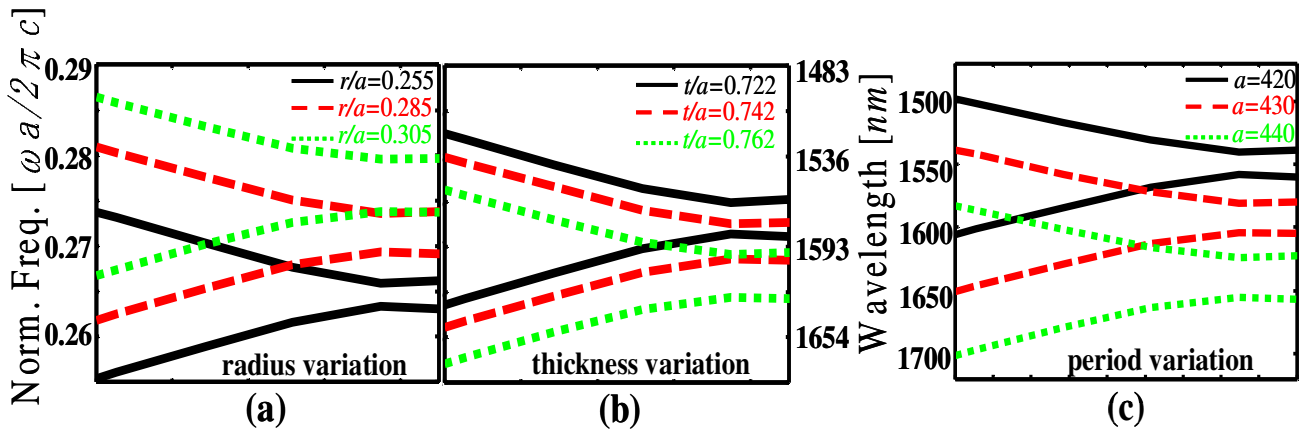

**Fig 2.** (a), Change of the band of the PhC with respect to radius (*r/a*) variation; *t/a* = 0.744. (b), Thickness (*t/a*) variation; *r/a* = 0.279. Wavelengths on the *y*-axis are calculated for *a* = 430 nm. (c) Period (*a*) variation. In numerical calculations, *t/a* and *r/a* are changed together so that effectively period variation is obtained.

Figure 2(a) shows the TM-like band diagram variation with *r*, where the bands shift to higher frequency with increasing *r/a* ratio and the band-gap broadens. Figure 2b shows the dependence of the photonic bands on *t*, the main conclusions being that the bands shift to lower frequency with increasing *t* and the band-gap broadens. Finally, we have varied *a*. In order to better see the corresponding variation of the photonic bands we plot in Fig. 2c our results versus wavelength instead of normalized frequency. As expected, the band diagram shifts to higher wavelength with increasing *a*.

The PhC slabs are integrated with MZI to facilitate the phase delay measurements as illustrated in Fig. 1a [12]. To examine the phase difference between designs containing different types of PhCs on the two arms and demonstrate experimentally optical filtering operation, we designed PhCs with the same number of unit cells but different geometrical parameters and kept *a* the same in order to have identical physical lengths on both arms [12]. By following this approach, all sections of the MZI, except PhC regions, are identical and therefore the two PhC sections are the only source for the measured phase difference.

The phase difference in the MZI is given by $\Delta\phi = (\beta_1 L_1 - \beta_2 L_2)$ where $\beta_i$ is the propagation constant and $L_i$ is the length of the MZI arms. Since $L_i$ is the same for the two arms, we have $\Delta\phi = \frac{2\pi}{\lambda}(n_1 - n_2)L$, where $n_i$ is the effective refractive index of the PhC sections on the arms.

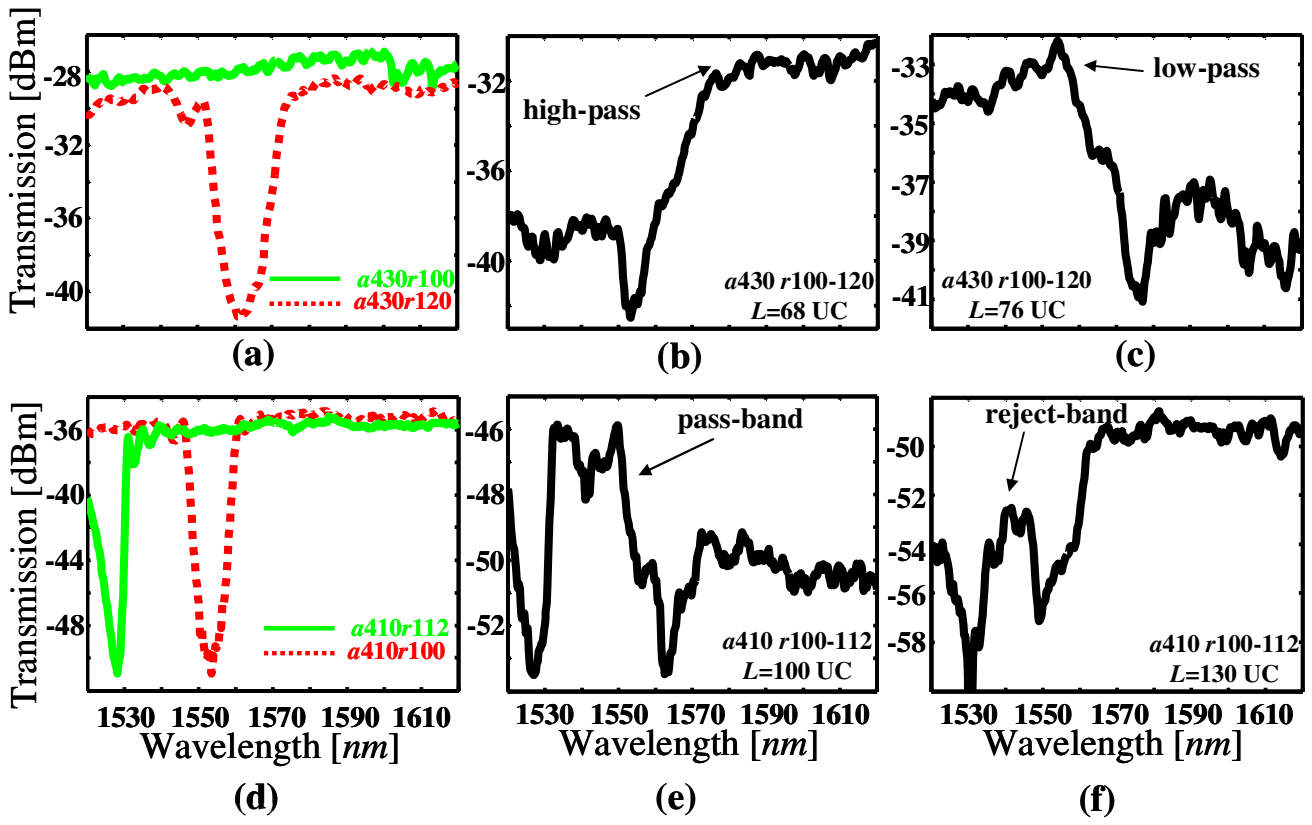

**Fig 3.** (a), Transmission spectrum for the designs on the two arms of the MZI in Figure 1a. (b), MZI transmission with the designs in (a) on the two arms with both 68 unit cells of PhC. (c), Same as in (b) but for 76 unit cells of PhC. (d), Same as in (a) with *a*=410nm. (e), MZI transmission with the designs in (d) on the two arms with both 100 unit cells of PhC. (f), Same as in (e) but for 130 unit cells of PhC.

Next, we fabricated [12] and tested our devices. Figure 3 presents experimental results demonstrating high-pass, low-pass, band-pass and band-reject filters. In Figure 3a, the transmission spectra for the two separate arms of the MZI are shown. Green (solid) line is the transmission of 80 unit cells of photonic crystal with *r* = 100 nm, *t* = 320 nm and *a* = 430 nm and the red (dashed) line is the transmission of 80 unit cells of PhC with *r* = 120 nm, *t* = 320 nm and *a* = 430 nm. In Figure 3b, the MZI transmission with the designs in Fig.3a (both having 68 unit cells) on the two arms is presented. The transmission is higher for the wavelengths between 1580 and 1620 nm and much lower for the wavelengths between 1520 and 1560 nm, a typical response of a high pass filter. Furthermore, when there are 76 unit cells of identical PhCs in both arms, the transmission is similar to a low pass filter.

Next, we changed the designs by decreasing the period so as to narrow the band-gap. In Fig. 3d, the transmission spectrum for individual designs with parameters *r* = 112 nm, *t* = 320 nm and *a* = 410 nm (solid line) and *r* = 100nm, *t* = 320 nm and *a* = 410 nm (dashed line) are shown. MZI transmission with these designs on the two arms of the MZI, with 100 unit cells of PhCs on each arm, is similar to that of a band-pass filter (the pass-band is between 1535 and 1550 nm, Fig 3e) and with 130 unit cells of PhC on each arm, a band-reject filter (see Fig. 3f).

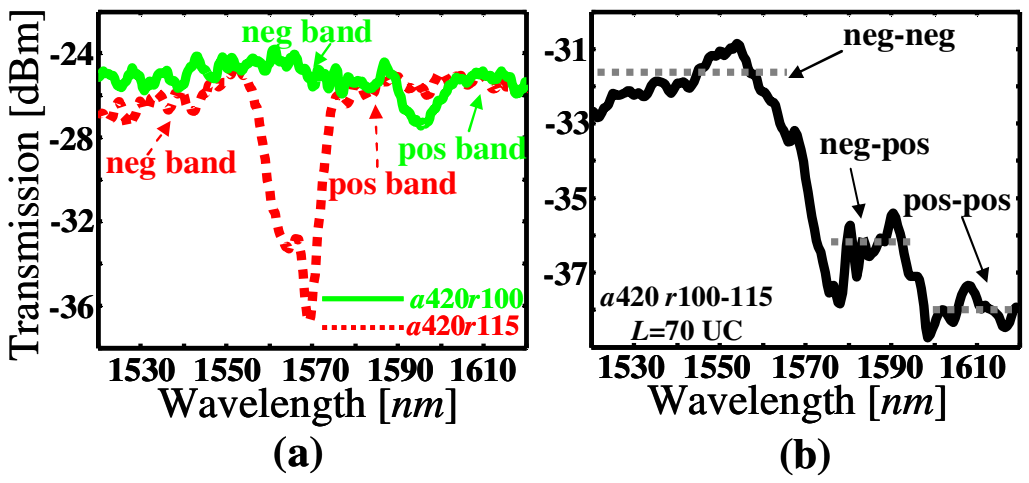

**Fig 4.** (a), Transmission spectrum for the designs on the two arms of the MZI in Figure 1a. (b), MZI transmission with the designs in (a) on the two arms with both 70 unit cells of PhC.

Finally, we designed a filter that has multi-level spectral response, namely, there are different rejection ratios for different spectral regions. Importantly, one can easily engineer both the width of the spectral regions as well as the rejection ratios. Again, we have analyzed the individual designs first. Figure 4a shows the transmission spectra for the PhCs inserted in the two arms of the MZI. The band-gap in the second design is not yet completely formed; however, this is not a critical issue as we operate at wavelengths at which the PhC is transparent. Figure 4b shows the MZI transmission with the designs presented in Fig. 4a and with 70 unit cells of PhC on both arms of the MZI. There are three transmission levels: first level is between 1520 nm and 1560 nm, where both PhCs have negative index values; second level is between 1575 nm and 1590 nm where the first PhC has positive index and the second one has negative index; and the third level is between 1600 nm and 1620 nm where both designs have positive index. This multi-level response filter is broadly tunable and can be used effectively in integrated photonics, *e.g.* as a variable attenuator.

In summary, we have demonstrated integrated optical filters that can be tuned in terms of bandwidth, operating wavelength, and rejection ratio *via* band diagram engineering of negative index PhC slabs. We observe filtering operations on CMOS-fabricated devices through intricate phase effects in the transparency region of PhCs instead of the spectral band-gap. Losses are mainly coming from coupling efficiency due to the lack of spot size converters. Importantly, rejection ratios can be improved by tuning the period *a* and using interferometer arms with different length as well as utilizing external phase tuning methods. These observations are the first demonstration of this type of chip-scale optical filters and can lead to exciting applications in photonic integrated circuits with scalable implementations.

The authors thank J.F. McMillan and J. Zheng for helpful discussions. We acknowledge funding support from NSF CAREER Award (0747787), NSF ECCS (0725707), and EPSRC (EP/G030502/1). Electron-beam nanopatterning was carried out at the Center for Functional Nanomaterials, Brookhaven National Laboratory, which is supported by the U.S. Department of Energy, Office of Basic Energy Sciences, under Contract No. DE-AC02-98CH10886.